\documentclass[prb,twocolumn,groupedaddress]{revtex4}

\usepackage{graphicx, bm, amsfonts, amssymb, amsmath, appendix}

\begin{document}

\title{Density-Dependent Analysis of Nonequilibrium Paths Improves Free Energy Estimates}

\author{David D. L. Minh}
\email[Electronic Address: ]{daveminh@gmail.com}

\affiliation{Laboratory of Chemical Physics, NIDDK, National Institutes of Health, Bethesda, Maryland 20892-0520, USA}

\date{\today}

\begin{abstract}
When a system is driven out of equilibrium by a time-dependent protocol that modifies the Hamiltonian, it follows a nonequilibrium path.  Samples of these paths can be used in nonequilibrium work theorems to estimate equilibrium quantities, such as free energy differences.  Here, we consider analyzing paths generated with one protocol using another one.  It is posited that analysis protocols which minimize the lag, the difference between the nonequilibrium and the instantaneous equilibrium densities, will reduce the dissipation of reprocessed trajectories and lead to better free energy estimates.  Indeed, when minimal lag analysis protocols based on exactly soluble propagators or relative entropies are applied to several test cases, substantial gains in the accuracy and precision of estimated free energy differences are observed.
\end{abstract}

\maketitle

\section{Introduction}

The accurate and efficient estimation of free energy differences is an important goal in chemical physics and remains an active area of research.  One promising approach to free energy estimation entails measuring the work done on a system over repetitions of an irreversible process.  According to the second law of thermodynamics, the mean work is greater than the free energy difference between the end states of the process, $F_\Lambda$.  Nonequilibrium work theorems \cite{Jarzynski1997a, Jarzynski1997b, Crooks1998, Crooks1999, Crooks2000} supplement this upper bound by rigorously equating $F_\Lambda$ with other averaged functions of the work.  These theorems have been empirically validated in single-molecule pulling experiments \cite{Liphardt2002, Collin2005} and computer simulations (e.g. Ref. \cite{Hummer2001b}).

Jarzynski's equality, \cite{Jarzynski1997a, Jarzynski1997b} a unidirectional nonequilibrium work theorem, relates the free energy difference to an exponential average of the work.  Unfortunately,  because it uses a nonlinear (specifically, a logarithmic) function of the average, the free energy estimator based on this equality suffers from a systematic finite-sampling bias. \cite{Zuckerman2002, Gore2003, Zuckerman2004}  While accurate in the limit of infinite sampling, this estimator is usually dominated by rare events where the work is less than the free energy difference, and thereby converges slowly. \cite{Jarzynski2006}

If the average amount of work dissipated as heat is reduced, these low-work events will be more frequent and accurate free energy estimation will usually require fewer work samples.  The most straightforward way to reduce heat dissipation is to slow the rate of the process; in the limit of infinitely slow switching, the process is reversible and the work is equal to the free energy difference.  Unfortunately, reducing the switching rate requires additional time and lowers the signal-to-noise ratio in single-molecule pulling experiments. \cite{Maragakis2008}  Under the constraint of constant experiment length, it is possible to reduce heat dissipation by optimizing the switching protocol that controls how the thermodynamic state changes with time.  Protocol variation predates Jarzynski's equality, having been applied to tightening free energy bounds from the second law of thermodynamics. \cite{Mark1990, Reinhardt1992, Hunter1993, Schon1996, Jarque1997}  More recently, variational calculus has been applied to find optimal protocols that minimize the mean work. \cite{Schmiedl2007, Then2008, GomezMarin2008}  

While protocol variation is, in principle, feasible in laboratory experiments, many more approaches to improving nonequilibrium-based free energy estimation are possible in computer simulations.  For example, Wu and Kofke were inspired by the Rosenbluth chain sampling scheme to develop methods for generating low-work nonequilibrium paths. \cite{Wu2005}  Vaikuntanathan and Jarzynski took another approach, altering the system dynamics, to reduce heat dissipation and improve free energy estimates.  \cite{Vaikuntanathan2008}  The approach most mathematically similar to this work, however, is importance sampling in nonequilibrium path space. \cite{Sun2003, Atilgan2004, Ytreberg2004, Oberhofer2005, Oberhofer2008}

In importance sampling, samples from one distribution are used to estimate expectations in another.  The technique is often applied to Markov chain Monte Carlo and molecular dynamics simulations (where it is usually called umbrella sampling): after applying a configurational bias to overcome energy barriers and promote ergodicity, expectations are calculated for the unbiased ensemble.  Importance sampling has been extended to transition path sampling \cite{Pratt1986, Dellago1998} with nonequilibrium trajectories.  In this algorithm, a biasing function modifies the Monte Carlo acceptance criteria of proposed paths in a way that improves the convergence of free energy estimates. \cite{Sun2003, Atilgan2004, Ytreberg2004, Oberhofer2005, Oberhofer2008}

Here, we apply the importance sampling formalism in a completely different way.  Instead of sampling nonequilibrium trajectories in a biased manner, we focus on the analysis of previously generated paths.  Instead of asking which path-ensemble we would like to sample from, we ask which path-ensemble average we would like to evaluate.  This is accomplished by processing paths generated using one protocol - the \emph{sampling} protocol, $\Lambda_s$ - using another - the \emph{analysis} protocol, $\Lambda$.

While we have infinite freedom in selecting an analysis protocol, not all choices will improve the convergence of free energy estimates.  One reasonable strategy for choosing $\Lambda$ is to minimize the lag, the difference between the nonequilibrium and instantaneous equilibrium densities; Vaikuntanathan and Jarzynski found that under certain dynamics, dissipation is eliminated if there is no lag, leading to a zero-variance estimator of $F_\Lambda$. \cite{Vaikuntanathan2008}  To reduce the lag, they modified their equation of motion with an additional flow-field term that ``escorts" the system along a near-equilibrium path.  Essentially, this strategy modifies the nonequilibrium density.  In this paper, we take the opposite approach: using the analysis protocol to choose an instantaneous equilibrium density that closely matches the sampled nonequilibrium density.

\begin{figure}
\begin{center}
\includegraphics{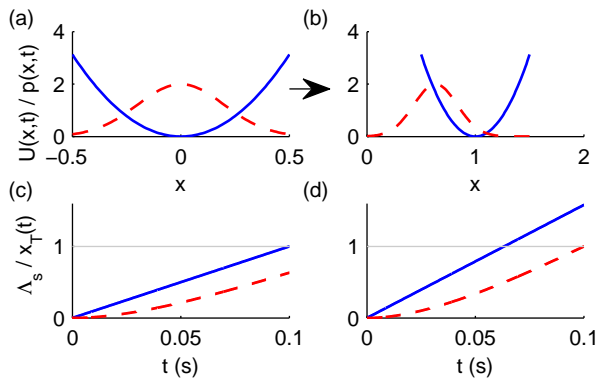}
\caption{ 
\label{fig:movingHO_lag}
Lag in a moving harmonic oscillator: Potential energy (solid line), $U(x,t)$, and density (dashed line), $p(x,t)$, as a function of position, at (a) $t = 0$ and (b) $t = 0.1$, where $v=10$.  Sampling protocol (solid line), $\Lambda_s$, and mean position, $x_T(t)$, as a function of time, for (c) $v = 10$ and (d) $v = 15.8019$.  For all parts of this figure, $D=1$ and $k=25$.
}
\end{center}
\end{figure}

As an illustrative case, consider a Brownian particle in a harmonic oscillator, or spring, which moves at a constant velocity (Fig.\ \ref{fig:movingHO_lag}).  If the system starts in thermal equilibrium, its density is a Gaussian about the initial spring position.  When the spring starts moving, the density remains a Gaussian with the same variance, but its mean position, $x_T(t)$, lags behind the spring position. \cite{Mazonka1999,Minh2009}  For this particular system, an analysis protocol based on $x_T(t)$ will have \emph{no} lag.  We shall further explore this system in Section \ref{sec:analytical}.

One complication with using an analysis protocol that minimizes the lag is that its end state is usually not the same as in the sampling protocol.  Thus, the free energy difference being estimated differs.  To estimate the same $F_\Lambda$ with a minimal lag analysis protocol, it may be necessary to extend or otherwise modify the sampling.  To distinguish the two situations, we shall refer to the former as \emph{protocol postprocessing} and the latter as \emph{nonequilibrium density-dependent sampling} (NEDDS).  Both fall under the aegis of density-dependent analysis.

The structure of this paper is as follows: in Section \ref{sec:FE}, the importance sampling form of Jarzynski's equality is detailed; in Section \ref{sec:analytical}, density-dependent analysis is demonstrated on two cases in which the propagator is analytically known; in Section \ref{sec:general}, a general method for finding minimal lag analysis protocols is described, applied to an adaptive algorithm for NEDDS, and tested on the model system; and lastly, implications of this method and possible future directions are discussed.

\section{\label{sec:FE}Free Energy Formalism}

Consider a system whose Hamiltonian, $H=H(x;\lambda)$, depends on $x$, its position in phase space (or configuration space), and a control parameter, $\lambda$.  Initially, the system is prepared in thermal equilibrium at $\lambda(0)$.  The parameter $\lambda$ is perturbed according to a protocol $\Lambda = \lambda(t)$ until it reaches a final state at $\lambda(\tau)$.  Jarznyski's equality, \cite{Jarzynski1997a, Jarzynski1997b}
\begin{eqnarray}
e^{-F_\Lambda} 
= \frac{ \int dX ~ e^{-W[X|\Lambda]} \rho_\Lambda[X] }{\int dX ~ \rho_\Lambda[X] }
\equiv \left< e^{-W[X|\Lambda]} \right>_\Lambda.
\label{eq:Jarz}
\end{eqnarray}
relates the free energy difference between the initial and final states of the protocol, $F_\Lambda$, to an average over all possible paths, $X = x(t)$, resulting from this nonequilibrium procedure.  Specifically, this expectation (denoted by the angled brackets $\left< ... \right>_{\Lambda}$), is a path integral over infinitesimal elements $dX$ with the protocol-dependent density $\rho_\Lambda[X]$.  During each process, the work done on the system is $W[X|\Lambda] = \int_0^{\tau} dt \dot{\lambda} ( \partial H / \partial \lambda)$.  (In this paper, all energies will be expressed in units of $k_B T$.)

Suppose that instead of $\rho_\Lambda[X]$, we consider an alternate density of paths, $\rho_s[X]$.  The free energy difference $F_\Lambda$ can be calculated by applying a reweighed form of Jarzynski's equality, \cite{Ytreberg2004}
\begin{eqnarray}
e^{-F_\Lambda}
= \frac{ \int dX ~ e^{-W[X|\Lambda]} \left( \frac{\rho_\Lambda[X]}{\rho_s[X]} \right) \rho_s[X] }
{\int dX ~ \left( \frac{\rho_\Lambda[X]}{\rho_s[X]} \right) \rho_s[X] }
\equiv \frac{ \left< r e^{-W[X|\Lambda]} \right>_s}{ \left< r \right>_s },
\label{eq:JarzISMP}
\end{eqnarray}
where $r = \rho_\Lambda[X] / \rho_s[X]$ is the ratio of probabilities of observing the trajectory given the densities.  To analyze a finite sample of paths drawn from $\rho_s[X]$, we replace the expectations with sample mean estimators, obtaining, \cite{Ytreberg2004}
\begin{equation}
\bar{F}_\Lambda = - \ln \left( \frac{ \sum_{n=1}^{N_s} r ~e^{- W[X_n | \Lambda]} }{ \sum_{n=1}^{N_s} r } \right),
\label{eq:JarzISMPsm}
\end{equation}
where $N_s$ is the sample size.  In a standard Jarzynski estimate, $r = 1$.

Previous workers have improved the convergence properties of Eq.\ (\ref{eq:JarzISMPsm}) by choosing $\rho_s[X]$ to be various work-weighted functionals of the original density $\rho_\Lambda[X]$.  \cite{Ytreberg2004, Oberhofer2005, Oberhofer2008}  When introducing the single-ensemble biased path sampling approach, Ytreberg and Zuckerman picked $\rho_s[X] = \rho_\Lambda[X] e^{-W[X|\Lambda]/2}$, such that $r = e^{W[X|\Lambda]/2}$. \cite{Ytreberg2004}  In a paper comparing the method with conventional equilibrium procedures, Oberhofer et. al. considered $\rho_s[X] = \rho_\Lambda[X] / P(W[X|\Lambda])$. \cite{Oberhofer2005}  By variation of the asymptotic variance with respect to the sampling bias, Oberhofer and Dellago found that optimal work-weighted sampling is given by $\rho_s[X] = \rho_\Lambda[x] |e^{-(W[X|\Lambda]-F_\Lambda)} - 1|$. \cite{Oberhofer2008}  Unfortunately, this optimal choice is impractical because it includes the sought quantity $F_\Lambda$.

In the present method, which applies Eq.\ (\ref{eq:JarzISMPsm}) in a novel manner, $\rho_s[X] = \rho_{\Lambda_s}[X]$ depends on the sampling protocol and $r$ differs from unity when the analysis protocol $\Lambda$ varies from $\Lambda_s$.  Notably, the relevant work is $W[X|\Lambda]$, not $W[X|\Lambda_s]$, meaning that different choices of $\Lambda$ will result in various work distributions and convergence properties.  This new way of applying importance sampling leads to different, albeit analogous, asymptotic variance expressions. \cite{Minh2009EPAPS}  The present approach is more general than previous applications of Eq.\ (\ref{eq:JarzISMPsm}), which require transition path sampling, because it does not require biased sampling and paths can be generated by ordinary dynamical equations.  Indeed, under certain assumptions, such as those suggested by Nummela and Andricioaei, \cite{Nummela2007} it should be possible to apply the present method to laboratory experiments.

We note, as a caveat, that the importance sampling form of Jarzynski's equality will only be useful for stochastic dynamics where $r$ can be computed.  Under deterministic dynamics, $r$ is a delta function and having different sampling and analysis protocols will not improve free energy estimates.

\section{\label{sec:analytical}Cases with an Analytical Propagator}

As mentioned earlier, we would like to choose an analysis protocol that minimizes the lag, such that the instantaneous equilibrium density corresponds with the sampled nonequilibrium density.  This is particularly tractable when the propagator is exactly known.  Here, we demonstrate Eq.\ (\ref{eq:JarzISMPsm}) on two such cases: a Brownian particle in a harmonic oscillator (i) moving at a constant velocity or (ii) with a time-dependent natural frequency.  With both, the potential energy has the general form $U(x) = k(x-\bar{x})^2/2$ and the nonequilibrium density is 
\begin{equation}
p_{neq}(x,t) = \sqrt{\frac{ k_T(t)}{2 \pi}} e^{-\frac{ k_T(t)}{2}(x-x_T(t))^2},
\label{eq:neq_density}
\end{equation}
where $x_T(t)$ and $k_T(t)$ are the most typical paths and spring coefficients, respectively.  As these propagators can be obtained by close analogy to the path integral derivation of \emph{work-weighted} propagators, \cite{Minh2009} their derivations are not detailed here.  In case (i), $k$ is constant and $\lambda$ moves the spring position according to $\bar{x} = \Lambda_s = vt$, such that $\Delta F = 0$, $k_T(t) = k$, and
\begin{equation}
x_T(t)  = vt - \frac{v}{D k} (1 - e^{-D kt}).
\label{eq:movingHO_x_T}
\end{equation}
In case (ii), $\bar{x}$ is zero and $\lambda$ controls the spring coefficient, $k = \Lambda_s$, such that $\Delta F = \frac{1}{2} \ln [k(0)/k(\tau)]$.  In the corresponding nonequilibrium density, $x_T(t) = 0$, and 
\begin{equation}
k_T(t) = \frac{ k(0) e^{2D \int_0^t ds~k(s)}}{1+2Dk(0)  \left[\int_0^t du~e^{2D \int_0^u ds~k(s)} \right]}.
\label{eq:changingHO_k_T}
\end{equation}
Based on these expressions, it is evident that, for case (i), the minimal lag analysis protocol is $\Lambda_{ml} = x_T(t)$ from Eq.\ (\ref{eq:movingHO_x_T}), and for case (ii), it is $\Lambda_{ml} = k_T(t)$ from Eq.\ (\ref{eq:changingHO_k_T}).  In these special cases, the nonequilibrium density is exactly the equilibrium density corresponding to $\Lambda_{ml}$ and there is \emph{no} lag.

To test whether density-dependent analysis leads to improved free energy estimates,  one-dimensional Brownian dynamics simulations were run with the equation of motion,
\begin{equation}
x_{j+1} = x_{j} - D \Delta t U_j' + \sqrt{2D\Delta t} R_j,
\label{eq:BD}
\end{equation}
where $x_j$ is the position at time $j \Delta t$, $\Delta t$ is the time step, and $R_j$ is a standard normal random variable.  The primes denote spatial derivatives such that $U_j' = \partial U(x_j;\lambda_j) / \partial x_j$ and $U_j'' = \partial^2 U(x_j;\lambda_j) / \partial x_j^2$.  For a discrete trajectory $X = \{x_0,x_1,...,x_J\}$ sampled with the protocol $\Lambda_s = \{\lambda_0, \lambda_1, ..., \lambda_J\}$, where $J$ is the total number of steps, the probability ratio is $r = e^{-\Delta S}$, where $\Delta S = S[X|\Lambda] - S[X|\Lambda_s]$ and $S[X|\Lambda]$ is the stochastic action (discretized from Ref. \cite{Minh2009}),
\begin{eqnarray}
S[X|\Lambda] & = & \frac{U(x_J;\lambda_J)+U(x_0;\lambda_0)}{2} \nonumber \\
& + & \frac{\Delta t}{4 D} \sum_{j=0}^{J-1} \left[ \left(\frac{x_{j+1} - x_{j}}{\Delta t} \right)^2 + (D U_j')^2 
- 2 D^2 U_j'' \right] \nonumber \\ 
& - & \frac{W[X|\Lambda]}{2}.
\end{eqnarray}
The work was evaluted with the discrete formula, $W[X|\Lambda] = \sum_{j=0}^{J-1} [U(x_{j+1};\lambda_{j+1}) - U(x_{j+1};\lambda_j)]$.  This action is valid in the continuum limit, as $J \rightarrow \infty$ and $\Delta t \rightarrow 0$.   To approach this limit, we chose $D = 1$ and a time step of $\Delta t = 0.001$.  

The simulations were performed over $10^{m}$ steps (truncated to be an integer), where m refers to 7 evenly spaced numbers between 1.5 and 3.  In case (i), $k$ was set to 25 and $\Lambda_s$ was chosen to start from $\lambda_0 = 0$ and linearly progress to the target state at $\lambda_f = 1$.  With case (ii), $\Lambda_s$ is a linear interpolation between 1 and 100.  Afterwards, the trajectories both analyzed with the standard Jarzynski estimate and subjected to protocol postprocessing with $\Lambda_{ml}$.

For comparison, NEDDS was implemented by switching $\lambda$ at a faster rate such that the final state went beyond $\lambda_f$ and the final nonequilibrium density, according to the propagators, corresponded to the target state.  This is illustrated in Fig.\ \ref{fig:movingHO_lag}d, where moving the harmonic oscillator at a faster rate than in Fig.\ \ref{fig:movingHO_lag}c allows for the final density to correspond to the equilibrium state with $\lambda = 1$.  These trajectories, which took the same amount of simulation time for the same number of steps, were then reanalyzed with the appropriate $\Lambda_{ml}$.

\begin{figure}
\begin{center}
\includegraphics{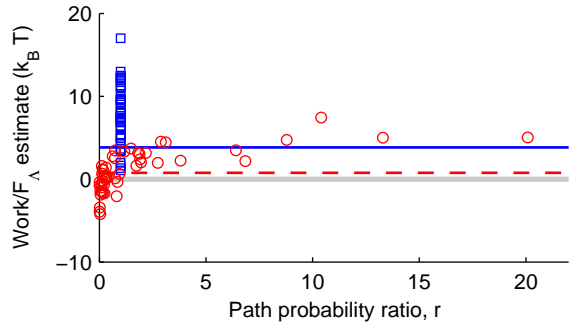}
\caption{\label{fig:movingHO_work_weight}
Representative work-weight plot for a moving harmonic oscillator: 
$W[X|\Lambda]$ and $r$ of 50 paths with $v=10$, analyzed with $\Lambda=\Lambda_s$ (squares) or $\Lambda=x_T(t)$ (circles).  The free energy difference (shaded line) and $\bar{F}_\Lambda$ from Eq.\ (\ref{eq:JarzISMPsm}) using $\Lambda=\Lambda_s$ (solid line) and $\Lambda=x_T(t)$ (dashed line) are denoted by horizontal lines.}
\end{center}
\end{figure}

In case (i), we find that protocol postprocessing with $\Lambda_{ml}$ leads to a desirable result: most work values are reduced such that a larger fraction of them are less than the free energy difference (Fig.\ \ref{fig:movingHO_work_weight}).  Of these negative dissipation trajectories, most have a probability ratio less than one.  Conversely, several positive dissipation trajectories have a probability ratio greater than one.  For this set of trajectories, the modified work distribution leads to a more accurate free energy estimate.

\begin{figure}
\begin{center}
\includegraphics{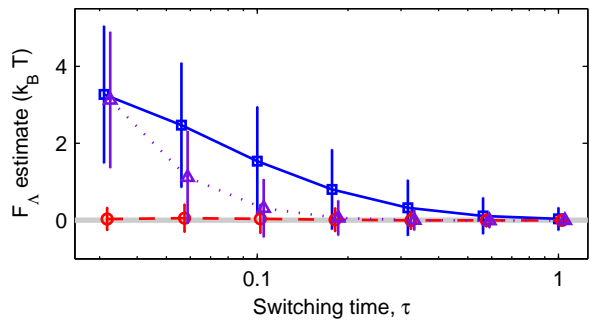}
\caption{\label{fig:movingHO_FE}
Comparison of free energy estimates for a moving harmonic oscillator:
Mean and standard deviation of 10000 $\bar{F}_\Lambda$ estimates using 50 trajectories each, analyzed with 
$\Lambda=\Lambda_s$ (squares), $\Lambda=\Lambda_{ml}$ (circles), or by NEDDS (triangles).  The latter two are slightly offset to prevent error bar overlap.}
\end{center}
\end{figure}

Over a large number of repetitions and range of switching speeds, we find that free energy estimates based on $\Lambda=\Lambda_{ml}$ are vastly improved over the standard procedure, $\Lambda = \Lambda_s$, having significantly less variance and systematic bias (Fig.\ \ref{fig:movingHO_FE}).  The standard estimator only approaches the accuracy and precision of protocol processing for slow switches.  Clearly, these trajectories are much better at estimating the end state free energy differences for $\Lambda_{ml}$ than for $\Lambda_s$.  The estimates of $F_\Lambda$ from NEDDS also require considerably less sampling than the standard procedure, although the effect is somewhat less dramatic.

\begin{figure}
\begin{center}
\includegraphics{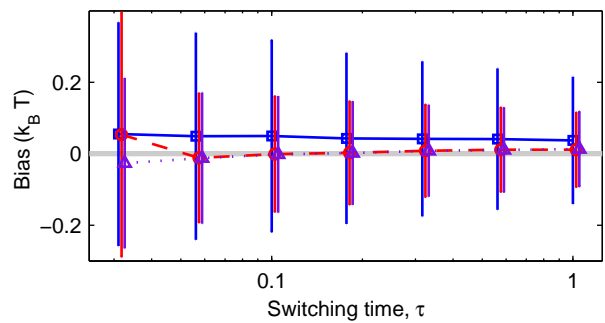}
\caption{\label{fig:changingHO_FE}
Comparison of free energy estimates for a harmonic oscillator with a time-dependent natural frequency:
Mean and standard deviation of 10000 $\bar{F}_\Lambda - F_\Lambda$ estimates using 50 trajectories each, analyzed with $\Lambda=\Lambda_s$ (squares), $\Lambda=k_T(t)$ (circles), or by NEDDS (triangles).  The latter two are slightly offset to prevent error bar overlap.}
\end{center}
\end{figure}

Similarly, in case (ii), density-dependent methods also show improvement over the standard Jarzynski estimate.  For the time-dependent natural frequency, the systematic bias of the standard estimate is relatively small but nonetheless evident at all sampled switching rates (Fig.\ \ref{fig:changingHO_FE}).  Estimates from both density-dependent methods have reduced bias and variance, and are found to be of similar quality to each other.

\section{\label{sec:general}General Case}

In most practical situations, unfortunately, the propagator is not known ahead of time.  Thus, prior to simulations, it is unclear how long paths need to be generated before the nonequilibrium density matches a density characteristic of the target state.  While paths are being generated, however, it is possible to estimate the difference between the sampled density and arbitrary equilibrium states.  States which minimize this difference can be be collected in an analysis protocol with minimal lag. 

One measure of the distance between two probability distributions is the Kullback-Leibler divergence, or the relative entropy.  The relative entropy between the nonequilibrium density and an arbitrary equilibrium state $T$ is,
\begin{eqnarray}
D_{KL}(p_{neq}(x,t) || p_T(x)) \equiv \int dx ~ p_{neq}(x,t) \ln \frac{p_{neq}(x,t)}{p_T(x)}.
\end{eqnarray}
When the integral is separated into two at the logarithm, one part is a constant with respect to $T$.  The divergence is minimized by finding a state where the other, $- \int dx ~ p_{neq}(x,t) \ln p_T(x)$, is the least.  Using sampled discrete paths, this integral can be estimated by $\sum_{n=1}^{N_s} \ln p_T(x_{jn})$, where $x_{jn}$ is the position at step $j$ of path $n$.  For a state $T$, the equilibrium density is $p_T(x) = \exp \left[-(H_T(x) - F_T(x)) \right]$, where $H_T(x)$ is the test state Hamiltonian and $F_T$ is its free energy.  Thus, the relative entropy is minimized by the smallest value of,
\begin{equation}
D_T(x_{j1},x_{j2},...,x_{jN_s}) = \frac{1}{N_s} \left[ \sum_{n=1}^{N_s} H_T(x_{jn}) \right] - F_T,
\label{eq:to_min_DKL}
\end{equation}
among different states $T$.  Generally, the free energy, $F_T$, is unknown, but for states which occur along the switching protocol, $F_T - F_0$ (where $F_0$ is the free energy at $\lambda_0$) can be estimated using the standard form of Jarzynski's equality.  These states constitute our search space for minimizing the lag.

Suppose we are interested in the free energy difference between the states defined by $\lambda_0$ and $\lambda_f$.  We can use $D_T$ to estimate $\Lambda_{ml}$ on the fly and determine when to stop sampling via the following adaptive algorithm:

\begin{enumerate}
\item Start with $j = 0$ and the work $W_0 = 0$.  For each of $N_s$ paths, obtain $x_0$ by drawing samples from the equilibrium ensemble at $\lambda_0$.
\item Propagate each path, calculating $x_{j+1}$ using a dynamical equation such as Eq.\ \ref{eq:BD}.  To obtain $W_{j+1}$, calculate the work done on the system during the time step and add it to $W_j$.  The next step in the sampling protocol, $\lambda_{j+1}$, is found by adding a predetermined value, $\mu$, to $\lambda_j$.  The sign of $\mu$ must be the same as $\lambda_f - \lambda_0$.  Increment $j$ by one.
\item Using $W_j$ values in the standard form of Jarzynski's equality, estimate $F_j - F_0$, the free energy difference between the states with $\lambda_j$ and $\lambda_0$.
\item For each state $T$ corresponding to $\{\lambda_0$, $\lambda_1$, ... $\lambda_j\}$, use $H_T(x)$ and the free energy difference estimated in the previous algorithm step to calculate $D_T - F_0$.  The $\lambda$ which minimizes $D_T - F_0$ is $\lambda_{ml}$.  Add $\lambda_{ml}$ to the minimal lag protocol $\Lambda_{ml}$.
\item If $\lambda_{ml}$ hasn't crossed $\lambda_f$, repeat from algorithm step 2.  Otherwise, set the final value in $\Lambda_{ml}$ to $\lambda_f$.
\item Estimate the free energy difference using Eq.\ (\ref{eq:JarzISMPsm}) with $\Lambda = \Lambda_{ml}$.
\end{enumerate}

This algorithm was tested on Sun's system, \cite{Sun2003} where the potential energy is $U(x) = x^4 - 16 \lambda x^2$.  Using Eq.\ (\ref{eq:JarzISMPsm}), the free energy difference was estimated between the initial state with $\lambda_0 = 0$, where the potential is a single well, and the target state $\lambda_f = 1$, where it is a double well, such that $\Delta F = -62.9407$. \cite{Oberhofer2005}  Brownian dynamics simulations were performed with the same diffusion coefficient, time step, and equation of motion as in Section \ref{sec:analytical}.  The increment of $\lambda$ at each time step was $\mu = v \Delta t$, where $v = 10^m$ and $m$ refers to 9 evenly spaced values between 0 and 2.  For comparison, the standard Jarzynski estimate was applied to simulations where $\lambda$ is switched between 0 and 1 at a slower velocity, taking the same total time as in the corresponding NEDDS simulations.

\begin{figure}
\begin{center}
\includegraphics{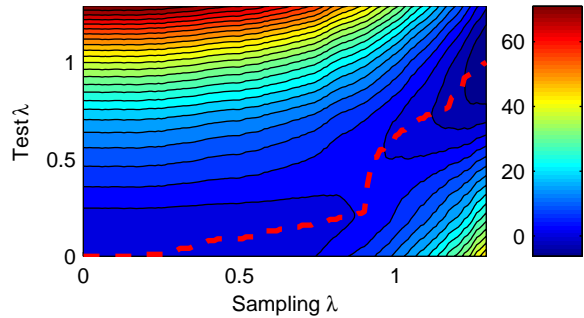}
\caption{\label{fig:SunF_div}
Representative divergence landscape for Sun's system: Contour plot of $D_T$ as a function of sampling $\lambda$, estimated using 50 paths with $v=10$.  $\Lambda_{ml}$ is shown with a dashed line.  Note that only half of this information, where the sampling $\lambda$ is less than the test $\lambda$, is available on-the-fly.
}
\end{center}
\end{figure}

In a representative set of simulations, the density most noticeably lags behind the sampling state at the beginning (Fig.\ \ref{fig:SunF_div}).  Around the state defined by $\lambda=0.9$, the lag quickly diminishes.  However, the minima of $D_T$ does not reach the target state until the sampling $\lambda$ is beyond 1.

\begin{figure}
\begin{center}
\includegraphics{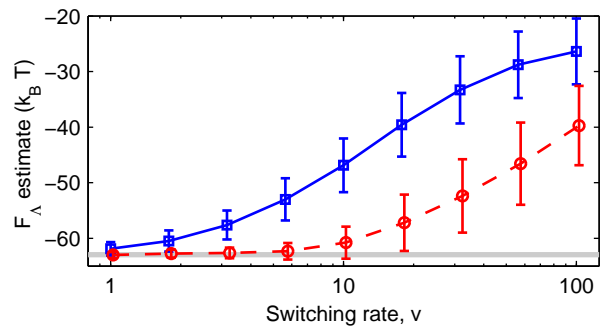}
\caption{\label{fig:SunF_FE}
Comparison of free energy estimates for Sun's system:
Mean and standard deviation of 10000 $\bar{F}_\Lambda$ estimates using 50 trajectories each, analyzed with 
$\Lambda=\Lambda_s$ (squares) or $\Lambda=\Lambda_{ml}$ (circles, slightly offset to prevent error bar overlap)
}
\end{center}
\end{figure}

Based on many repetitions of this procedure at different pulling speeds, we find that our NEDDS algorithm converges much more quickly than the standard Jarzynski estimate (Fig.\ \ref{fig:SunF_FE}).  The systematic bias is largely eliminated with simulations that are switched nearly an order of magnitude faster.  At the fastest switching rates, NEDDS remains biased but still outperforms the standard Jarzynski estimate.  With these fast switchings, it is possible that the nonequilibrium density does not correspond well to \emph{any} traversed equilibrium state.

\section{Discussion and Conclusion}

With the goal of minimizing the lag via the choice of analysis protocol, we have developed density-dependent methods to analyze nonequilbrium paths, to estimate which states may constitute a protocol that minimizes the lag, and to adaptively sample paths until the desired density is achieved.  Our promising results validate the strategy and provide further evidence for the link between lag and heat dissipation.  They also hint that the accurate estimation of free energy differences may require adequate sampling in the important regions of both end states.

Analysis protocols provide another degree of freedom for lag reduction, and can be used in conjunction with other methods, such as sampling protocol optimization or biased path sampling.  Furthermore, their use should extend beyond Jarzynski's equality; they can potentially be applied in bidirectional nonequilibrium work expressions \cite{Minh2009EPAPS} or any relationship between a nonequilibrium process and a state function, such as Hummer and Szabo's expression for the potential of mean force. \cite{Hummer2001a}  Quite possibly, our results are just the tip of an iceberg and this paper will open up new research directions for sampling and analyzing nonequilibrium trajectories.

\section{Acknowledgments}

The author thanks Artur Adib, Christopher Jarzynski, Attila Szabo, and Suriyanarayanan Vaikuntanathan for pertinent discussions, and Gerhard Hummer for suggesting that he considers the lag.  He also thanks Artur Adib for supporting a postdoctoral fellowship.  This research was supported by the Intramural Research Program of the NIH, NIDDK.


\begin{thebibliography}{36}
\expandafter\ifx\csname natexlab\endcsname\relax\def\natexlab#1{#1}\fi
\expandafter\ifx\csname bibnamefont\endcsname\relax
  \def\bibnamefont#1{#1}\fi
\expandafter\ifx\csname bibfnamefont\endcsname\relax
  \def\bibfnamefont#1{#1}\fi
\expandafter\ifx\csname citenamefont\endcsname\relax
  \def\citenamefont#1{#1}\fi
\expandafter\ifx\csname url\endcsname\relax
  \def\url#1{\texttt{#1}}\fi
\expandafter\ifx\csname urlprefix\endcsname\relax\def\urlprefix{URL }\fi
\providecommand{\bibinfo}[2]{#2}
\providecommand{\eprint}[2][]{\url{#2}}

\bibitem[{\citenamefont{Jarzynski}(1997{\natexlab{a}})}]{Jarzynski1997a}
\bibinfo{author}{\bibfnamefont{C.}~\bibnamefont{Jarzynski}},
  \bibinfo{journal}{Phys. Rev. Lett.} \textbf{\bibinfo{volume}{78}},
  \bibinfo{pages}{2690} (\bibinfo{year}{1997}{\natexlab{a}}).

\bibitem[{\citenamefont{Jarzynski}(1997{\natexlab{b}})}]{Jarzynski1997b}
\bibinfo{author}{\bibfnamefont{C.}~\bibnamefont{Jarzynski}},
  \bibinfo{journal}{Phys. Rev. E} \textbf{\bibinfo{volume}{56}},
  \bibinfo{pages}{5018} (\bibinfo{year}{1997}{\natexlab{b}}).

\bibitem[{\citenamefont{Crooks}(1998)}]{Crooks1998}
\bibinfo{author}{\bibfnamefont{G.~E.} \bibnamefont{Crooks}},
  \bibinfo{journal}{J. Stat. Phys.} \textbf{\bibinfo{volume}{90}},
  \bibinfo{pages}{1481} (\bibinfo{year}{1998}).

\bibitem[{\citenamefont{Crooks}(1999)}]{Crooks1999}
\bibinfo{author}{\bibfnamefont{G.~E.} \bibnamefont{Crooks}},
  \bibinfo{journal}{Phys. Rev. E} \textbf{\bibinfo{volume}{60}},
  \bibinfo{pages}{2721} (\bibinfo{year}{1999}).

\bibitem[{\citenamefont{Crooks}(2000)}]{Crooks2000}
\bibinfo{author}{\bibfnamefont{G.~E.} \bibnamefont{Crooks}},
  \bibinfo{journal}{Phys. Rev. E} \textbf{\bibinfo{volume}{61}},
  \bibinfo{pages}{2361} (\bibinfo{year}{2000}).

\bibitem[{\citenamefont{Liphardt et~al.}(2002)\citenamefont{Liphardt, Dumont,
  Smith, {Tinoco~Jr.}, and Bustamante}}]{Liphardt2002}
\bibinfo{author}{\bibfnamefont{J.}~\bibnamefont{Liphardt}},
  \bibinfo{author}{\bibfnamefont{S.}~\bibnamefont{Dumont}},
  \bibinfo{author}{\bibfnamefont{S.~B.} \bibnamefont{Smith}},
  \bibinfo{author}{\bibfnamefont{I.}~\bibnamefont{{Tinoco~Jr.}}},
  \bibnamefont{and}
  \bibinfo{author}{\bibfnamefont{C.}~\bibnamefont{Bustamante}},
  \bibinfo{journal}{Science} \textbf{\bibinfo{volume}{296}},
  \bibinfo{pages}{1832} (\bibinfo{year}{2002}).

\bibitem[{\citenamefont{Collin et~al.}(2005)\citenamefont{Collin, Ritort,
  Jarzynski, Smith, Tinoco, and Bustamante}}]{Collin2005}
\bibinfo{author}{\bibfnamefont{D.}~\bibnamefont{Collin}},
  \bibinfo{author}{\bibfnamefont{F.}~\bibnamefont{Ritort}},
  \bibinfo{author}{\bibfnamefont{C.}~\bibnamefont{Jarzynski}},
  \bibinfo{author}{\bibfnamefont{S.~B.} \bibnamefont{Smith}},
  \bibinfo{author}{\bibfnamefont{I.}~\bibnamefont{Tinoco}}, \bibnamefont{and}
  \bibinfo{author}{\bibfnamefont{C.}~\bibnamefont{Bustamante}},
  \bibinfo{journal}{Nature} \textbf{\bibinfo{volume}{437}},
  \bibinfo{pages}{231} (\bibinfo{year}{2005}).

\bibitem[{\citenamefont{Hummer}(2001)}]{Hummer2001b}
\bibinfo{author}{\bibfnamefont{G.}~\bibnamefont{Hummer}}, \bibinfo{journal}{J.
  Chem. Phys.} \textbf{\bibinfo{volume}{114}}, \bibinfo{pages}{7330}
  (\bibinfo{year}{2001}).

\bibitem[{\citenamefont{Zuckerman and Woolf}(2002)}]{Zuckerman2002}
\bibinfo{author}{\bibfnamefont{D.~M.} \bibnamefont{Zuckerman}}
  \bibnamefont{and} \bibinfo{author}{\bibfnamefont{T.~B.} \bibnamefont{Woolf}},
  \bibinfo{journal}{Phys. Rev. Lett.} \textbf{\bibinfo{volume}{89}},
  \bibinfo{pages}{180602} (\bibinfo{year}{2002}).

\bibitem[{\citenamefont{Gore et~al.}(2003)\citenamefont{Gore, Ritort, and
  Bustamante}}]{Gore2003}
\bibinfo{author}{\bibfnamefont{J.}~\bibnamefont{Gore}},
  \bibinfo{author}{\bibfnamefont{F.}~\bibnamefont{Ritort}}, \bibnamefont{and}
  \bibinfo{author}{\bibfnamefont{C.}~\bibnamefont{Bustamante}},
  \bibinfo{journal}{Proc. Natl. Acad. Sci. U.S.A.}
  \textbf{\bibinfo{volume}{100}}, \bibinfo{pages}{12564}
  (\bibinfo{year}{2003}).

\bibitem[{\citenamefont{Zuckerman and Woolf}(2004)}]{Zuckerman2004}
\bibinfo{author}{\bibfnamefont{D.~M.} \bibnamefont{Zuckerman}}
  \bibnamefont{and} \bibinfo{author}{\bibfnamefont{T.~B.} \bibnamefont{Woolf}},
  \bibinfo{journal}{J. Stat. Phys.} \textbf{\bibinfo{volume}{114}},
  \bibinfo{pages}{1303} (\bibinfo{year}{2004}).

\bibitem[{\citenamefont{Jarzynski}(2006)}]{Jarzynski2006}
\bibinfo{author}{\bibfnamefont{C.}~\bibnamefont{Jarzynski}},
  \bibinfo{journal}{Phys. Rev. E} \textbf{\bibinfo{volume}{73}},
  \bibinfo{pages}{046105} (\bibinfo{year}{2006}).

\bibitem[{\citenamefont{Maragakis et~al.}(2008)\citenamefont{Maragakis, Ritort,
  Bustamante, Karplus, and Crooks}}]{Maragakis2008}
\bibinfo{author}{\bibfnamefont{P.}~\bibnamefont{Maragakis}},
  \bibinfo{author}{\bibfnamefont{F.}~\bibnamefont{Ritort}},
  \bibinfo{author}{\bibfnamefont{C.}~\bibnamefont{Bustamante}},
  \bibinfo{author}{\bibfnamefont{M.}~\bibnamefont{Karplus}}, \bibnamefont{and}
  \bibinfo{author}{\bibfnamefont{G.~E.} \bibnamefont{Crooks}},
  \bibinfo{journal}{J. Chem. Phys.} \textbf{\bibinfo{volume}{129}},
  \bibinfo{pages}{024102} (\bibinfo{year}{2008}).

\bibitem[{\citenamefont{Mark et~al.}(1990)\citenamefont{Mark, {van~Gunsteren},
  and Berendsen}}]{Mark1990}
\bibinfo{author}{\bibfnamefont{A.~E.} \bibnamefont{Mark}},
  \bibinfo{author}{\bibfnamefont{W.~F.} \bibnamefont{{van~Gunsteren}}},
  \bibnamefont{and} \bibinfo{author}{\bibfnamefont{H.~J.~C.}
  \bibnamefont{Berendsen}}, \textbf{\bibinfo{volume}{94}},
  \bibinfo{pages}{3808} (\bibinfo{year}{1990}).

\bibitem[{\citenamefont{Reinhardt and Hunter}(1992)}]{Reinhardt1992}
\bibinfo{author}{\bibfnamefont{W.~P.} \bibnamefont{Reinhardt}}
  \bibnamefont{and} \bibinfo{author}{\bibfnamefont{J.~E.}
  \bibnamefont{Hunter}}, \bibinfo{journal}{J. Chem. Phys.}
  \textbf{\bibinfo{volume}{97}}, \bibinfo{pages}{1599} (\bibinfo{year}{1992}).

\bibitem[{\citenamefont{Hunter et~al.}(1993)\citenamefont{Hunter, Reinhardt,
  and Davis}}]{Hunter1993}
\bibinfo{author}{\bibfnamefont{J.~E.} \bibnamefont{Hunter}},
  \bibinfo{author}{\bibfnamefont{W.~P.} \bibnamefont{Reinhardt}},
  \bibnamefont{and} \bibinfo{author}{\bibfnamefont{T.~F.} \bibnamefont{Davis}},
  \bibinfo{journal}{J. Chem. Phys.} \textbf{\bibinfo{volume}{99}},
  \bibinfo{pages}{6856} (\bibinfo{year}{1993}).

\bibitem[{\citenamefont{Schon}(1996)}]{Schon1996}
\bibinfo{author}{\bibfnamefont{J.~C.} \bibnamefont{Schon}},
  \bibinfo{journal}{J. Chem. Phys.} \textbf{\bibinfo{volume}{105}},
  \bibinfo{pages}{10072} (\bibinfo{year}{1996}).

\bibitem[{\citenamefont{Jarque and Tidor}(1997)}]{Jarque1997}
\bibinfo{author}{\bibfnamefont{C.}~\bibnamefont{Jarque}} \bibnamefont{and}
  \bibinfo{author}{\bibfnamefont{B.}~\bibnamefont{Tidor}}, \bibinfo{journal}{J.
  Phys. Chem. B} \textbf{\bibinfo{volume}{101}}, \bibinfo{pages}{9402}
  (\bibinfo{year}{1997}).

\bibitem[{\citenamefont{Schmiedl and Seifert}(2007)}]{Schmiedl2007}
\bibinfo{author}{\bibfnamefont{T.}~\bibnamefont{Schmiedl}} \bibnamefont{and}
  \bibinfo{author}{\bibfnamefont{U.}~\bibnamefont{Seifert}},
  \bibinfo{journal}{Phys. Rev. Lett.} \textbf{\bibinfo{volume}{98}},
  \bibinfo{pages}{108301} (\bibinfo{year}{2007}).

\bibitem[{\citenamefont{Then and Engel}(2008)}]{Then2008}
\bibinfo{author}{\bibfnamefont{H.}~\bibnamefont{Then}} \bibnamefont{and}
  \bibinfo{author}{\bibfnamefont{A.}~\bibnamefont{Engel}},
  \bibinfo{journal}{Phys. Rev. E} \textbf{\bibinfo{volume}{77}},
  \bibinfo{pages}{041105} (\bibinfo{year}{2008}).

\bibitem[{\citenamefont{Gomez-Marin et~al.}(2008)\citenamefont{Gomez-Marin,
  Schmiedl, and Seifert}}]{GomezMarin2008}
\bibinfo{author}{\bibfnamefont{A.}~\bibnamefont{Gomez-Marin}},
  \bibinfo{author}{\bibfnamefont{T.}~\bibnamefont{Schmiedl}}, \bibnamefont{and}
  \bibinfo{author}{\bibfnamefont{U.}~\bibnamefont{Seifert}},
  \bibinfo{journal}{J. Chem. Phys.} \textbf{\bibinfo{volume}{129}},
  \bibinfo{pages}{024114} (\bibinfo{year}{2008}).

\bibitem[{\citenamefont{Wu and Kofke}(2005)}]{Wu2005}
\bibinfo{author}{\bibfnamefont{D.}~\bibnamefont{Wu}} \bibnamefont{and}
  \bibinfo{author}{\bibfnamefont{D.~A.} \bibnamefont{Kofke}},
  \bibinfo{journal}{J. Chem. Phys.} \textbf{\bibinfo{volume}{122}},
  \bibinfo{eid}{204104} (\bibinfo{year}{2005}).

\bibitem[{\citenamefont{Vaikuntanathan and
  Jarzynski}(2008)}]{Vaikuntanathan2008}
\bibinfo{author}{\bibfnamefont{S.}~\bibnamefont{Vaikuntanathan}}
  \bibnamefont{and}
  \bibinfo{author}{\bibfnamefont{C.}~\bibnamefont{Jarzynski}},
  \bibinfo{journal}{Phys. Rev. Lett.} \textbf{\bibinfo{volume}{100}},
  \bibinfo{pages}{190601} (\bibinfo{year}{2008}).

\bibitem[{\citenamefont{Sun}(2003)}]{Sun2003}
\bibinfo{author}{\bibfnamefont{S.}~\bibnamefont{Sun}}, \bibinfo{journal}{J.
  Chem. Phys.} \textbf{\bibinfo{volume}{118}}, \bibinfo{pages}{5769}
  (\bibinfo{year}{2003}).

\bibitem[{\citenamefont{Atilgan and Sun}(2004)}]{Atilgan2004}
\bibinfo{author}{\bibfnamefont{E.}~\bibnamefont{Atilgan}} \bibnamefont{and}
  \bibinfo{author}{\bibfnamefont{S.~X.} \bibnamefont{Sun}},
  \bibinfo{journal}{J. Chem. Phys.} \textbf{\bibinfo{volume}{121}},
  \bibinfo{pages}{10392} (\bibinfo{year}{2004}).

\bibitem[{\citenamefont{Ytreberg and Zuckerman}(2004)}]{Ytreberg2004}
\bibinfo{author}{\bibfnamefont{F.~M.} \bibnamefont{Ytreberg}} \bibnamefont{and}
  \bibinfo{author}{\bibfnamefont{D.~M.} \bibnamefont{Zuckerman}},
  \bibinfo{journal}{J. Chem. Phys.} \textbf{\bibinfo{volume}{120}},
  \bibinfo{pages}{10876} (\bibinfo{year}{2004}).

\bibitem[{\citenamefont{Oberhofer et~al.}(2005)\citenamefont{Oberhofer,
  Dellago, and Geissler}}]{Oberhofer2005}
\bibinfo{author}{\bibfnamefont{H.}~\bibnamefont{Oberhofer}},
  \bibinfo{author}{\bibfnamefont{C.}~\bibnamefont{Dellago}}, \bibnamefont{and}
  \bibinfo{author}{\bibfnamefont{P.}~\bibnamefont{Geissler}},
  \bibinfo{journal}{J. Phys. Chem. B} \textbf{\bibinfo{volume}{109}},
  \bibinfo{pages}{6902} (\bibinfo{year}{2005}).

\bibitem[{\citenamefont{Oberhofer and Dellago}(2008)}]{Oberhofer2008}
\bibinfo{author}{\bibfnamefont{H.}~\bibnamefont{Oberhofer}} \bibnamefont{and}
  \bibinfo{author}{\bibfnamefont{C.}~\bibnamefont{Dellago}},
  \bibinfo{journal}{Comput. Phys. Commun.} \textbf{\bibinfo{volume}{179}},
  \bibinfo{pages}{41} (\bibinfo{year}{2008}).

\bibitem[{\citenamefont{Pratt}(1986)}]{Pratt1986}
\bibinfo{author}{\bibfnamefont{L.}~\bibnamefont{Pratt}}, \bibinfo{journal}{J.
  Chem. Phys.} \textbf{\bibinfo{volume}{85}}, \bibinfo{pages}{5045}
  (\bibinfo{year}{1986}).

\bibitem[{\citenamefont{Dellago et~al.}(1998)\citenamefont{Dellago, Bolhuis,
  Csajka, and Chandler}}]{Dellago1998}
\bibinfo{author}{\bibfnamefont{C.}~\bibnamefont{Dellago}},
  \bibinfo{author}{\bibfnamefont{P.~G.} \bibnamefont{Bolhuis}},
  \bibinfo{author}{\bibfnamefont{F.~S.} \bibnamefont{Csajka}},
  \bibnamefont{and} \bibinfo{author}{\bibfnamefont{D.}~\bibnamefont{Chandler}},
  \bibinfo{journal}{J. Chem. Phys.} \textbf{\bibinfo{volume}{108}},
  \bibinfo{pages}{1964} (\bibinfo{year}{1998}).

\bibitem[{\citenamefont{Mazonka and Jarzynski}(1999)}]{Mazonka1999}
\bibinfo{author}{\bibfnamefont{O.}~\bibnamefont{Mazonka}} \bibnamefont{and}
  \bibinfo{author}{\bibfnamefont{C.}~\bibnamefont{Jarzynski}},
  \emph{\bibinfo{title}{Exactly solvable model illustrating
  far-from-equilibrium predictions}} (\bibinfo{year}{1999}),
  \eprint{cond-mat/9912121}.

\bibitem[{\citenamefont{Minh and Adib}(2009)}]{Minh2009}
\bibinfo{author}{\bibfnamefont{D.~D.~L.} \bibnamefont{Minh}} \bibnamefont{and}
  \bibinfo{author}{\bibfnamefont{A.~B.} \bibnamefont{Adib}},
  \bibinfo{journal}{Phys. Rev. E} \textbf{\bibinfo{volume}{79}},
  \bibinfo{pages}{021122} (\bibinfo{year}{2009}).

\bibitem[{Min()}]{Minh2009EPAPS}
\bibinfo{note}{See appendices}.

\bibitem[{\citenamefont{Nummela and Andricioaei}(2007)}]{Nummela2007}
\bibinfo{author}{\bibfnamefont{J.}~\bibnamefont{Nummela}} \bibnamefont{and}
  \bibinfo{author}{\bibfnamefont{I.}~\bibnamefont{Andricioaei}},
  \bibinfo{journal}{Biophys. J.} \textbf{\bibinfo{volume}{93}},
  \bibinfo{pages}{3373} (\bibinfo{year}{2007}).

\bibitem[{\citenamefont{Hummer and Szabo}(2001)}]{Hummer2001a}
\bibinfo{author}{\bibfnamefont{G.}~\bibnamefont{Hummer}} \bibnamefont{and}
  \bibinfo{author}{\bibfnamefont{A.}~\bibnamefont{Szabo}},
  \bibinfo{journal}{Proc. Natl. Acad. Sci. U.S.A.}
  \textbf{\bibinfo{volume}{98}}, \bibinfo{pages}{3658} (\bibinfo{year}{2001}).

\bibitem[{\citenamefont{Shirts et~al.}(2003)\citenamefont{Shirts, Bair, Hooker,
  and Pande}}]{Shirts2003}
\bibinfo{author}{\bibfnamefont{M.~R.} \bibnamefont{Shirts}},
  \bibinfo{author}{\bibfnamefont{E.}~\bibnamefont{Bair}},
  \bibinfo{author}{\bibfnamefont{G.}~\bibnamefont{Hooker}}, \bibnamefont{and}
  \bibinfo{author}{\bibfnamefont{V.~S.} \bibnamefont{Pande}},
  \bibinfo{journal}{Phys. Rev. Lett.} \textbf{\bibinfo{volume}{91}},
  \bibinfo{pages}{140601} (\bibinfo{year}{2003}).

\end{thebibliography}

\newpage

\begin{widetext}
\begin{appendix}

In these appendices, we derive asymptotic variance and bias expressions for free energies estimated using protocol postprocessing.  Our derivations are similar to that of Oberhofer et. al. for biased sampling of nonequilibrium trajectories. \cite{Oberhofer2005, Oberhofer2008}  Although these expressions are not directly relevant to the main text, they support the line of inquiry explored in the paper and may be useful to those who wish to expand upon the results.  Both unidirectional and bidirectional expressions are considered.


\section{Unidirectional}

Expressed in importance sampling form, Jarzynski's equality is,
\begin{eqnarray}
e^{- F_\Lambda} =
\frac{\left< r~e^{- W[X|\Lambda]} \right>_{s} }
{\left< r \right>_{s} },
\label{eq:JarzPP2}
\end{eqnarray}
This is Eq.\ (3) in the main text, reproduced here for convenience.

Towards deriving the asymptotic expressions, we first define,
\begin{eqnarray}
A & = & r~e^{- W[X|\Lambda]} \\
B & = & r
\end{eqnarray}
The sample mean estimators for the expectations of $A$ and $B$ are,
\begin{eqnarray}
\bar A = \frac{1}{N_s} \sum_{n=1}^{N_s} A_n \\
\bar B = \frac{1}{N_s} \sum_{n=1}^{N_s} B_n
\end{eqnarray}
These form an estimator for the free energy,
\begin{equation}
\bar{F}_\Lambda = - \ln (\bar A / \bar B)
\label{eq:free_energy_AB}
\end{equation}
However, the sample mean estimators deviate from their true expectations,
\begin{eqnarray}
\bar A = \left< A \right> - \Delta \bar{A} \\
\bar B = \left< B \right> - \Delta \bar{B}
\end{eqnarray}
Here, subscripts on the angle brackets are omitted for notational simplicity.

Deviations in these expectations lead to variance and bias in $\bar{F}_\Lambda$.  The magnitude of the error can be estimated by a Taylor series expansion about $F_\Lambda$, which becomes an increasingly reasonable approximation in the large sampling, or asymptotic, limit.  To the first order, this expansion is,
\begin{eqnarray}
\bar F_\Lambda & = & - \ln \frac{\left< A \right> - \Delta \bar{A}}{\left< B \right> - \Delta \bar{B}} \\
& \approx & F_\Lambda
+ \left( \frac{\partial \bar F_\Lambda}{\partial \bar{A}} \right) \Delta \bar{A}
+ \left( \frac{\partial \bar F_\Lambda}{\partial \bar{B}} \right) \Delta \bar{B} \\
& = & F_\Lambda -  \left( \frac{\Delta \bar{A}}{\left< A \right>} - \frac{\Delta \bar{B}}{\left< B \right>} \right)
\end{eqnarray}
The partial derivatives are evaluated at their mean values.

The variance is defined as $\sigma^2[\bar{F}_\Lambda] \equiv \left< (\bar{F}_\Lambda - F_\Lambda)^2 \right>$.  Using the first-order Taylor series expansion, this is,
\begin{eqnarray}
 \sigma^2[\bar{F}_\Lambda] & \approx & \left< \left( \frac{\Delta \bar{A}}{\left< A \right>} - \frac{\Delta \bar{B}}{\left< B \right>} \right)^2 \right> \\
& = & \left< \frac{\Delta \bar{A}^2}{\left< A \right>^2} + \frac{\Delta \bar{B}^2}{\left< B \right>^2} 
- 2 \frac{\Delta \bar{A}\Delta \bar{B}}{\left< A \right> \left< B \right>} \right> \\
& = & \frac{\left< \Delta \bar{A}^2 \right>}{\left< A \right>^2} + \frac{\left< \Delta \bar{B}^2 \right>}{\left< B \right>^2} 
- 2 \frac{\left< \Delta \bar{A}\Delta \bar{B} \right>}{\left< A \right> \left< B \right>} 
\label{eq:var_sample_mean}
\\
& = & \frac{\sigma^2[\bar{A}]}{\left< A \right>^2} + \frac{\sigma^2[\bar{B}]}{\left< B \right>^2} 
- 2 \frac{\sigma^2[\bar{A},\bar{B}]}{\left< A \right> \left< B \right>}
\end{eqnarray}

The variance of a sample mean is the variance of the variable over the number of samples.  Since $A$ and $B$ are functions of the same data points, the covariance of their sample means is, similarly, $\sigma^2[\bar{A},\bar{B}] = \frac{1}{N} \sigma^2[A,B]$.  Thus, the asymptotic variance is,
\begin{eqnarray}
\sigma^2[\bar{F}_\Lambda] & = & \frac{1}{N} \left[ \frac{\left< A^2 \right>}{\left< A \right>^2} 
+ \frac{\left< B^2 \right>}{\left< B \right>^2} - \frac{2 \left< A B \right>}{\left< A \right> \left< B \right>} \right] \\
& = &
\frac{1}{N} \left[ \frac{ \left< A^2 \right> e^{2  F_\Lambda} + \left< B^2 \right> - 2 \left< AB \right> e^{  F_\Lambda}}{ \left< B \right>^2} \right] \\
& = & 
\frac{1}{N_s} \frac{ \left< r^2 e^{-2  (W[X|\Lambda] - F_\Lambda)} + r^2 - 2 r^2 e^{- (W[X|\Lambda] - F_\Lambda)} \right>_{s} }
{ \left< r \right>_{s}^2} \\
& = & 
\frac{1}{N_s} \frac{ \left< r^2 (e^{- (W[X|\Lambda] - F_\Lambda)} - 1)^2 \right>_{s} }{ \left< r \right>_{s}^2}
\label{eq:var_JarzPP}
\end{eqnarray}

The bias is defined as $B_N \equiv \left< \bar F_\Lambda \right> - F_\Lambda$.  If we approximate this error with a first-order Taylor series expansion, it is always zero.  In order to obtain a nonzero bias expression, we use a second-order Taylor series expansion about $F_\Lambda$,
\begin{eqnarray}
\bar F_\Lambda & = & - \ln \frac{\left< A \right> - \Delta \bar{A}}{\left< B \right> - \Delta \bar{B}} \\
& \approx & F_\Lambda
+ \frac{\partial \bar F_\Lambda}{\partial \bar{A}} \Delta \bar{A}
+ \frac{\partial \bar F_\Lambda}{\partial \bar{B}} \Delta \bar{B}
+ \frac{1}{2} \frac{\partial^2 \bar F_\Lambda}{\partial \bar{A}^2} \Delta \bar{A}^2 
+ \frac{\partial^2 \bar F_\Lambda}{\partial \bar{A} \partial \bar{B}} \Delta \bar{A} \Delta \bar{B} 
+ \frac{1}{2} \frac{\partial^2 \bar F_\Lambda}{\partial \bar{B}^2} \Delta \bar{B}^2 \\
& = & F_\Lambda - 
\left( 
\frac{\Delta \bar{A}}{\left< A \right>} - \frac{\Delta \bar{B}}{\left< B \right>} 
- \frac{1}{2} \frac{\Delta \bar{A}^2}{\left< A \right>^2} 
+ \frac{1}{2} \frac{\Delta \bar{B}^2}{\left< B \right>^2} 
\right)
\end{eqnarray}

Using this approximation, the bias is found to be,
\begin{eqnarray}
B_N & \approx & \frac{1}{2 } \left[ \frac{ \left< \Delta \bar{A}^2 \right> }{\left< A \right>^2} - \frac{ \left< \Delta \bar{B}^2 \right> }{\left< B \right>^2}
 \right] \\
& = & \frac{1}{2  N} \left[ \frac{ \left< A^2 \right> }{ \left< A \right>^2 } - \frac{ \left< B^2 \right> }{ \left< B \right>^2 } \right] \\
& = & \frac{1}{2  N} \left[ \frac{ \left< A^2 \right> e^{2  F_\Lambda} }{ \left< B \right>^2 } - \frac{ \left< B^2 \right> }{ \left< B \right>^2 } \right] \\
& = & \frac{1}{2  N_s} \frac{\left< r^2 \left( e^{-2  (W[X|\Lambda] - F_\Lambda)} - 1 \right) \right>_{s}}
{\left< r \right>_{s}^2} 
\end{eqnarray}

Notably, when the dissipated work is zero, $W[X|\Lambda] - F_\Lambda = 0$, expressions for both the variance and bias are likewise zero.


\section{Bidirectional}

Here, we consider the possibility of reanalyzing bidirectional data, collected using both a protocol $\Lambda$ and its time reversal $\tilde{\Lambda}$.  The results derived in this section suggest that for bidirectional data, the optimal analysis protocol is actually the sampling protocol.  Thus, protocol postprocessing is less promising when applied to bidirectional data than to unidirectional data.

For notational consistency, we start our discussion with the Crooks Fluctuation Theorem,\cite{Crooks1998, Crooks1999}
\begin{equation}
\frac{\rho_{\Lambda}[X]}{\rho_{\tilde{\Lambda}}[\tilde{X}]} = e^{W[X|\Lambda] - F_\Lambda}
\label{eq:CFT}
\end{equation}
As in the main text, $\rho_\Lambda[X]$ is the probability of observing trajectory $X$, given the protocol $\Lambda$.  Analogously, $\rho_{\tilde{\Lambda}}[\tilde{X}]$ is the probability of observing the time reversal, or conjugate twin, of $X$, using the reverse protocol $\tilde{\Lambda}$.

This theorem can be used to derive a relationship between forward and reverse path-ensemble averages, \cite{Crooks2000}
\begin{eqnarray}
\left< \mathcal F_\Lambda[X] \right>_\Lambda & = & \int dX ~ \mathcal F_\Lambda[X] \rho_\Lambda[X] \\
& = & \int d \tilde{X} ~ \mathcal F_\Lambda[X] e^{W[X|\Lambda] - F_\Lambda} \rho_{\tilde{\Lambda}}[\tilde{X}] \\
& = & \left< \mathcal F_\Lambda[X]  e^{-W[\tilde{X}|\tilde{\Lambda}] - F_\Lambda} \right>_{\tilde{\Lambda}}
\label{eq:CPATH}
\end{eqnarray}
In the above, $\mathcal F_\Lambda[X]$ is an arbitrary functional.  As the last path-ensemble average is over trajectories $\tilde{X}$, trajectories sampled from $\tilde{\Lambda}$ must be reversed prior to being evaluated with the functional.

Next, we rearrange the path-ensemble average theorem into an expression for the free energy. \cite{Crooks2000}
\begin{equation}
e^{- F_\Lambda} = 
\frac{ \left< \mathcal F_\Lambda[X] \right>_\Lambda }
{ \left< \mathcal F_\Lambda[X]  e^{- W[\tilde{X}|\tilde{\Lambda}]} \right>_{\tilde{\Lambda}} }
\label{eq:CPATH_dF}
\end{equation}
As in the unidirectional case, these path-ensemble averages can be written as reweighed samples from other sampling densities.
\begin{eqnarray}
e^{- F_\Lambda} & = &
\frac{ \left< r \mathcal F_\Lambda[X] \right>_{s} }{ \left< r \right>_{s} }
\frac{ \left< \tilde{r} \right>_{\tilde{s}} }
{ \left< \tilde{r} \mathcal F_\Lambda[X]  e^{- W[\tilde{X}|\tilde{\Lambda}]} \right>_{\tilde{s}} } \\
& = &
\frac{ \left< r \mathcal F_\Lambda[X] \right>_{s} }
{ \left< \tilde{r} \mathcal F_\Lambda[X]  e^{- W[\tilde{X}|\tilde{\Lambda}]} \right>_{\tilde{s}} }
\label{eq:biPPunoptimized}
\end{eqnarray}
The probability ratio $\tilde{r}$ is defined similarly to $r$,
\begin{equation}
\tilde{r} = \frac{ \rho_{\tilde{\Lambda}}[\tilde{X}] }{ \rho_{\tilde{s}}[\tilde{X}] }
\end{equation}
Using Eq.\ (\ref{eq:CFT}), we can show that it is related to $r$ by
\begin{equation}
r = \tilde{r} e^{W[X|\Lambda] - W[X|\Lambda_s]}
\label{eq:r_and_rhat}
\end{equation}
The ratio of $\left< \tilde{r} \right>_{\tilde{s}}/ \left< r \right>_{s}$ can be shown to be unity by converting $\left< \tilde{r} \right>_{\tilde{s}}$ into a forward path-ensemble average using Eqs.~(\ref{eq:CPATH}) and (\ref{eq:r_and_rhat}), and applying the importance sampling form of Jarzynski's equality, Eq.\ (\ref{eq:JarzPP2}).

The asymptotic variance of Eq.\ (\ref{eq:biPPunoptimized}) can be calculated by a similar procedure to the unidirectional case.  We start by defining,
\begin{eqnarray}
C & = & r \mathcal F_\Lambda[X] \\
D & = & \tilde{r} \mathcal F_\Lambda[X] e^{- W[\tilde{X}|\tilde{\Lambda]}}
\end{eqnarray}
Replacing $A$ and $B$ with $C$ and $D$, we follow the same logic as in the unidirectional case from Eqs.~(\ref{eq:free_energy_AB}) to (\ref{eq:var_sample_mean}).  Next, we note that C and D are independent samples drawn from different ensembles and their correlation is zero.  Thus, the variance estimate is,
\begin{eqnarray}
\sigma^2[\bar{F}_\Lambda] & = & 
\frac{\left< C^2 \right>}{N_s \left< C \right>^2} 
+ \frac{\left< D^2 \right>}{N_{\tilde{s}} \left< D \right>^2} - \left(\frac{1}{N_s} + \frac{1}{N_{\tilde{s}}}\right) \\
& = &
\frac{\left< r^2 \mathcal F^2_\Lambda[X] \right>_{s}}{N_s \left< r \mathcal F_\Lambda[X] \right>_{s}^2} 
+ \frac{\left<  \tilde{r}^2 \mathcal F^2_\Lambda[X] e^{-2  W[\tilde{X}|\tilde{\Lambda]}} \right>_{\tilde{s}}}{N_{\tilde{s}} \left<  \tilde{r} \mathcal F_\Lambda[X] e^{- W[\tilde{X}|\tilde{\Lambda]}} \right>_{\tilde{s}}^2} - \left(\frac{1}{N_s} + \frac{1}{N_{\tilde{s}}}\right)
\label{eq:var_biPP}
\end{eqnarray}

We would like to combine the two terms including C and D in a single path-ensemble average.  In order to do that we need to convert the ensemble averages containing D to the forward direction.  For $\left< D \right>$, this is,
\begin{eqnarray}
\left< D \right>
& = & \left<  \tilde{r} \mathcal F_\Lambda[X] e^{- W[\tilde{X}|\tilde{\Lambda]}} \right>_{\tilde{s}} \\
& = & \int d\tilde{X} ~ \tilde{r} ~ \mathcal F_\Lambda[X] 
e^{- W[\tilde{X}|\tilde{\Lambda]}} ~ \rho_{\tilde{s}}[\tilde{X}] \\
& = & \int dX ~ r e^{W[X|\Lambda_s]-W[X|\Lambda]} ~ \mathcal F_\Lambda[X] 
e^{ W[X|\Lambda]} ~ \rho_s[X] e^{-W[X|\Lambda_s] + F_\Lambda} \\
& = & \int dX ~ r \mathcal F_\Lambda[X] \rho_s[X] e^{ F_\Lambda} \\
& = & \left< r \mathcal F_\Lambda[X] \right>_{s} e^{ F_\Lambda}
\label{eq:meanD}
\end{eqnarray}

For $\left< D^2 \right>$, this is
\begin{eqnarray}
\left< D^2 \right>
& = & \left<  \tilde{r}^2 \mathcal F^2_\Lambda[X] e^{-2  W[\tilde{X}|\tilde{\Lambda]}} \right>_{\tilde{s}} \\
& = & \int d\tilde{X} ~ \tilde{r}^2 \mathcal ~ F^2_\Lambda[X] 
e^{-2  W[\tilde{X}|\tilde{\Lambda]}} ~ \rho_{\tilde{s}}[\tilde{X}] \\
& = & \int dX ~ r^2 e^{2(W[X|\Lambda_s]-W[X|\Lambda])} ~ \mathcal F^2_\Lambda[X] 
e^{2 W[X|\Lambda]} ~ \rho_s[X] e^{-W[X|\Lambda_s] + F_\Lambda} \\
& = & \int dX ~ r^2 \mathcal F^2_\Lambda[X] 
e^{ W[X|\Lambda_s]} \rho_s[X] e^{ F_\Lambda} \\
& = & \left< r^2 \mathcal F^2_\Lambda[X] e^{ W[X|\Lambda_s]} \right>_{s} e^{ F_\Lambda}
\label{eq:meanD2}
\end{eqnarray}

Using Eqs.~(\ref{eq:var_biPP}),(\ref{eq:meanD}), and (\ref{eq:meanD2}), we obtain,
\begin{equation}
\sigma^2[\bar{F}_\Lambda] = 
\frac{ \left< r^2 \mathcal F^2_\Lambda[X] \left[ \frac{1}{N_s} + \frac{1}{N_{\tilde{s}}} e^{W[X|\Lambda_s] - F_\Lambda} \right] \right>_{s}}
{ \left< r \mathcal F_\Lambda[X] \right>_{s}^2 } - \left(\frac{1}{N_s} + \frac{1}{N_{\tilde{s}}}\right)
\label{eq:var_biPP_unoptimized}
\end{equation}

Now if we choose the functionals,
\begin{eqnarray}
\mathcal F_\Lambda [X] & = & r^{-1} \left[ \frac{1}{N_s} + \frac{1}{N_{\tilde{s}}} e^{W[X|\Lambda_s] - F_\Lambda} \right]^{-1} \\
\mathcal F_{\tilde{\Lambda}} [\tilde{X}] & = & \tilde{r}^{-1}
e^{W[\tilde{X}|\tilde{\Lambda}] - W[\tilde{X}|\tilde{\Lambda}_s]}
\left[ \frac{1}{N_s} + \frac{1}{N_{\tilde{s}}} e^{-W[\tilde{X}|\tilde{\Lambda}_s] - F_\Lambda} \right]^{-1}
\end{eqnarray}
then we obtain a generalized form of the Bennett Acceptance Ratio, as derived by Crooks. \cite{Crooks2000}

Variational optimization of Eq.\ (\ref{eq:var_biPP_unoptimized}), however, leads to the functionals,
\begin{eqnarray}
\mathcal F_\Lambda [X] & = & r^{-2} \left[ \frac{1}{N_s} + \frac{1}{N_{\tilde{s}}} e^{W[X|\Lambda_s] - F_\Lambda} \right]^{-1} \\
\mathcal F_{\tilde{\Lambda}} [\tilde{X}] & = & \tilde{r}^{-2}
e^{2 (W[\tilde{X}|\tilde{\Lambda}] - W[\tilde{X}|\tilde{\Lambda}_s])}
\left[ \frac{1}{N_s} + \frac{1}{N_{\tilde{s}}} e^{-W[\tilde{X}|\tilde{\Lambda}_s] - F_\Lambda} \right]^{-1}
\end{eqnarray}

Substituting these into Eq.\ (\ref{eq:biPPunoptimized}) leads to, 
\begin{eqnarray}
e^{- F_\Lambda} & = &
\frac{ \left< \frac{1}{ r \left[ \frac{1}{N_s} + \frac{1}{N_{\tilde{s}}} e^{W[X|\Lambda_s] - F_\Lambda} \right]} \right>_{s} }
{ \left< \frac{e^{W[\tilde{X}|\tilde{\Lambda}] - 2 W[\tilde{X}|\tilde{\Lambda}_s]}}
{\tilde{r} \left[ \frac{1}{N_s} + \frac{1}{N_{\tilde{s}}} e^{-W[\tilde{X}|\tilde{\Lambda}_s] - F_\Lambda} \right] } \right>_{\tilde{s}} } \\
& = & \frac{ N_s \left< \frac{1}{ r \left[ 1 + e^{M + W[X|\Lambda_s] - F_\Lambda} \right]} \right>_{s} }
{ N_{\tilde{s}} \left< \frac{e^{W[\tilde{X}|\tilde{\Lambda}] - W[\tilde{X}|\tilde{\Lambda}_s]}
e^{-W[\tilde{X}|\tilde{\Lambda}_s] - F_\Lambda} }
{\tilde{r} \left[ e^{- M} + e^{-W[\tilde{X}|\tilde{\Lambda}_s] - F_\Lambda} \right] }\right>_{\tilde{s}}e^{ F_\Lambda} } \\
& = &
\frac{ \left< \frac{1}{ r \left[ 1 + e^{M + W[X|\Lambda_s] - F_\Lambda} \right]} \right>_{s} }
{ \left< \frac{e^{W[\tilde{X}|\tilde{\Lambda}] - W[\tilde{X}|\tilde{\Lambda}_s]} }
{\tilde{r} \left[ e^{- (M - W[\tilde{X}|\tilde{\Lambda}_s] - F_\Lambda) } + 1 \right] }\right>_{\tilde{s}} } e^{ M - F_\Lambda}
\end{eqnarray}
where $M =  \ln \frac{N_s}{N_{\tilde{s}}}$.

This expression is analogous to Bennett's original expression.  It can be solved self-consistently or by rearrangement into,
\begin{equation}
N_s \left< \frac{1}{ r \left[ 1 + e^{M + W[X|\Lambda_s] - F_\Lambda} \right]} \right>_{s} - 
N_{\tilde{s}} \left< \frac{e^{W[\tilde{X}|\tilde{\Lambda}] - W[\tilde{X}|\tilde{\Lambda}_s]} }
{\tilde{r} \left[ 1 + e^{- (M - W[\tilde{X}|\tilde{\Lambda}_s] - F_\Lambda) } \right] }\right>_{\tilde{s}} = 0
\end{equation}
Using the sample mean estimator for the expectations, this is,
\begin{equation}
\sum_{n=1}^{N_s} \frac{1}{ r \left[ 1 + e^{M + W[X_n|\Lambda_s] - F_\Lambda} \right] } - \sum_{j=1}^{N_{\tilde{s}}} \frac{e^{W[\tilde{X}_j|\tilde{\Lambda}] - W[\tilde{X}_j|\tilde{\Lambda}_s]} }
{\tilde{r} \left[ 1 + e^{- (M - W[\tilde{X}_j|\tilde{\Lambda}_s] - F_\Lambda) } \right] } = 0
\label{eq:biPP}
\end{equation}
which is similar to the expression of Shirts et. al. for the Bennett Acceptance Ratio \cite{Shirts2003}.  This is an implicit function of $F_\Lambda$ which is solved by finding the zero of the equation.

The variance of this expression can be found by plugging the optimal functionals into Eq.\ (\ref{eq:var_biPP_unoptimized}),
\begin{equation}
\sigma^2[\bar{F}_\Lambda] = 
\frac{ \left< r^{-2} \left[ \frac{1}{N_s} + \frac{1}{N_{\tilde{s}}} e^{W[X|\Lambda_s] - F_\Lambda} \right]^{-1} \right>_{s}}
{ \left< r^{-1} \left[ \frac{1}{N_s} + \frac{1}{N_{\tilde{s}}} e^{W[X|\Lambda_s] - F_\Lambda} \right]^{-1} \right>_{s}^2 } - \left(\frac{1}{N_s} + \frac{1}{N_{\tilde{s}}}\right)
\label{eq:var_biPP_optimized1}
\end{equation}

We would like to express this equation in a form in which it is clear how to include data sampled from forward and reverse path-ensembles.  The path-ensemble average in the numerator is,
\begin{eqnarray}
\label{eq:eform}
\left< r^{-2} \left[ \frac{1}{N_s} + \frac{1}{N_{\tilde{s}}} e^{W[X|\Lambda_s] - F_\Lambda} \right]^{-1} \right>_{s} & = &
\int dX ~ \frac{N_s \rho_s[X]}{r^{2} \left[1 + e^{M + W[X|\Lambda_s] - F_\Lambda} \right]}  \\
\label{eq:coshform}
& = & \int dX ~ \frac{N_s \rho_s[X] + 
N_{\tilde{s}} \rho_{\tilde{s}}[\tilde{X}]}
{r^2 \left[ 2 + 2 \cosh ( M + W[X|\Lambda_s] - F_\Lambda ) \right]}
\end{eqnarray}
We obtain Eq.\ (\ref{eq:coshform}) from Eq.\ (\ref{eq:eform}) by multiplying it by 
\begin{equation}
\frac{ 1 + e^{-(M + W[X|\Lambda_s] - F_\Lambda)} }{ 1 + e^{-(M + W[X|\Lambda_s] - F_\Lambda)} }.
\end{equation}
Splitting the integral into two, we obtain a form amenable to treating forward and reverse switching data,
\begin{equation}
\int dX ~ \frac{N_s \rho_s[X]}
{r^2 \left[ 2 + 2 \cosh ( M + W[X|\Lambda_s] - F_\Lambda) \right]}
+ \int d\tilde{X} ~ \frac{N_{\tilde{s}} \rho_{\tilde{s}}[\tilde{X}] 
e^{2 (W[\tilde{X}|\tilde{\Lambda}] - W[\tilde{X}|\tilde{\Lambda}_s])} }
{\tilde{r}^2 \left[ 2 + 2 \cosh ( M - W[\tilde{X}|\tilde{\Lambda}_s] - F_\Lambda ) \right]}
\end{equation}
which can be estimated with,
\begin{equation}
\sum_{n=1}^{N_s} \frac{1}
{r^2 \left[ 2 + 2 \cosh ( M + W[X_n|\Lambda_s] - F_\Lambda ) \right]}
+ \sum_{j=1}^{N_{\tilde{s}}} \frac{e^{2 (W[\tilde{X}_j|\tilde{\Lambda}] - W[\tilde{X}_j|\tilde{\Lambda}_s])} }
{\tilde{r}^2 \left[ 2 + 2 \cosh ( M - W[\tilde{X}_j|\tilde{\Lambda}_s] - F_\Lambda ) \right]}
\end{equation}
By analogous procedure, the path-ensemble average in the denominator of Eq.\ (\ref{eq:var_biPP_optimized1}) is,
\begin{equation}
\sum_{n=1}^{N_s} \frac{1}
{r \left[ 2 + 2 \cosh (M + W[X_n|\Lambda_s] - F_\Lambda) \right]}
+ \sum_{j=1}^{N_{\tilde{s}}} \frac{e^{W[\tilde{X}_j|\tilde{\Lambda}] - W[\tilde{X}_j|\tilde{\Lambda}_s]} }
{\tilde{r} \left[ 2 + 2 \cosh (M - W[\tilde{X}_j|\tilde{\Lambda}_s] - F_\Lambda) \right]}
\end{equation}
Finally, we obtain the variance estimator of the bidirectional free energy calculation,
\begin{equation}
\sigma^2[\bar{F}_\Lambda] = 
\frac{ \sum_{n=1}^{N_s} \frac{1}
{r^2 \left[ 2 + 2 \cosh (M + W[X_n|\Lambda_s] - F_\Lambda) \right]}
+ \sum_{j=1}^{N_{\tilde{s}}} \frac{e^{2 (W[\tilde{X}_j|\tilde{\Lambda}] - W[\tilde{X}_j|\tilde{\Lambda}_s])} }
{\tilde{r}^2 \left[ 2 + 2 \cosh (M - W[\tilde{X}_j|\tilde{\Lambda}_s] - F_\Lambda) \right]} }
{\left( 
\sum_{n=1}^{N_s} \frac{1}
{r \left[ 2 + 2 \cosh (M + W[X_n|\Lambda_s] - F_\Lambda) \right]}
+ \sum_{j=1}^{N_{\tilde{s}}} \frac{e^{W[\tilde{X}_j|\tilde{\Lambda}] - W[\tilde{X}_j|\tilde{\Lambda}_s]} }
{\tilde{r} \left[ 2 + 2 \cosh (M - W[\tilde{X}_j|\tilde{\Lambda}_s] - F_\Lambda) \right]}
\right)^2}
- \left(\frac{1}{N_s} + \frac{1}{N_{\tilde{s}}} \right)
\end{equation}

As with unidirectional data, a natural question to ask is how to find the optimal analysis protocol when processing bidirectional data.  It no longer makes sense to reduce the lag; a protocol which reduces the lag for forward trajectories may increase it for trajectories from the reverse protocol.  Furthermore, when Eq.\ (\ref{eq:var_biPP_optimized1}) is variationally optimized with respect to $r$, the optimal $r$ is found to be constant.  Since this only occurs when $\Lambda = \Lambda_s$, this result suggests that, compared with unidirectional data, protocol processing less likely to be useful for treating bidirectional data.

\end{appendix}
\end{widetext}

\end{document}